\documentclass[twocolumn,showpacs,preprintnumbers,amsmath,amssymb]{revtex4}

\usepackage{graphicx}

\newcommand{\sm}{\, {\rm M}_{\odot}}

\newcommand{\kms}{km s$^{-1}$}

\begin{document}


\title[The phase-space structure of a dark-matter halo]{The
phase-space structure of a dark-matter halo: \\ Implications for
dark-matter direct detection experiments}


\author{Amina Helmi, Simon D.M. White and Volker Springel}
\email{ahelmi, swhite, volker@mpa-garching.mpg.de}
\affiliation{Max-Planck-Institut f\"ur Astrophysik, Karl-Schwarzschild-Str.
1, 85740 Garching bei M\"unchen, Germany} 



\date{\today}

\begin{abstract}
We study the phase-space structure of a dark-matter halo formed in a
high resolution simulation of a $\Lambda$CDM cosmology.  Our goal is
to quantify how much substructure is left over from the inhomogeneous
growth of the halo, and how it may affect the signal in experiments
aimed at detecting the dark matter particles directly.  If we focus on
the equivalent of ``Solar vicinity'', we find that the dark-matter is
smoothly distributed in space. The probability of detecting particles
bound within dense lumps of individual mass less than $10^7 M_\odot
h^{-1}$ is small, less than $10^{-2}$. The velocity
ellipsoid in the Solar neighbourhood deviates only slightly from a
multivariate Gaussian, and can be thought of as a superposition of
thousands of kinematically cold streams. The motions of the most
energetic particles are, however, strongly clumped and highly
anisotropic. We conclude that experiments may safely assume a smooth
multivariate Gaussian distribution to represent the kinematics of
dark-matter particles in the Solar neighbourhood.  Experiments
sensitive to the direction of motion of the incident particles could
exploit the expected anisotropy to learn about the recent merging
history of our Galaxy.
\end{abstract}

\maketitle

\section{Introduction}

One of the most fundamental open questions in cosmology and particle
physics today is what is the nature of dark-matter. The first
indications of its existence came in the 1930s, with the measurements
of the velocities of galaxies in clusters. The cluster mass required
to gravitationally bind the galaxies was found to be roughly an order
of magnitude larger than the sum of the luminous masses of the
individual galaxies \cite{zwicky,smith}.  In the 1970s, observations
of the rotation curves of spiral galaxies ($V_c(r) = \sqrt{GM(r)/r}$)
showed that these were flat or even rising at distances far beyond
their stellar and gaseous components
\cite{rubin70,faber79,rubin80}. These discoveries led to the
conclusion that a large fraction (more than 90\%) of the mass in the
Universe is dark. It is now widely believed that this mass is most
likely in the form of yet to be discovered nonbaryonic elementary
particles.

Being the dominant mass component of galaxies and of large-scale
structures in the Universe, dark-matter has necessarily become a key
ingredient in theories of structure formation in the Universe. The
most successful of these theories is the hierarchical paradigm
\cite{peebles74}. In the current (and observationally most favoured)
version of this model, the nonbaryonic elementary particles are known
as ``cold dark-matter" (CDM) \cite{peebles82}.  The term ``cold"
derives from the fact that the dark-matter particles had
non-relativistic motions at the time of matter--radiation
equality. Their abundance was set when the interaction rate became too
small for the particles to be in thermal equilibrium in the expanding
Universe. 

The first objects to form in a CDM Universe are small galaxies, which
then merge and give rise to the larger scale structures we observe
today. Thus structure formation occurs in a ``bottom-up" fashion
{\cite{blumenthal84,frenk83}.  This hierarchical paradigm has allowed
astronomers to make very definite predictions for the properties of
galaxies today and about their evolution from high redshift. Direct
comparisons to observations have shown that this model is quite
successful in reproducing both the local and the distant Universe.

The crucial test of this paradigm undoubtedly consists in the
determination of the nature of dark-matter through direct detection
experiments. Among the most promising candidates from the particle
physics perspective are axions and neutralinos.  Axions have been
introduced to solve the strong-CP (Charge conjugation and Parity)
violations \cite{PC}. They can be detected through their conversion to
photons in the presence of a strong magnetic field (e.g.
\cite{ax1,ax2}). Neutralinos are the lightest supersymmetric
particles, and may be considered as a particular form of weakly
interacting massive particles (WIMPs). The most important direct detection
process of neutralinos is through elastic scattering on nuclei. The
idea is to determine the count rate over recoil energy above a given
(detector) background level. The experimental situation has been
improving rapidly over the past years, with large-scale collaborations
such as DAMA, Edelweiss and CDMS ~\cite{DAMA00,CDMS00,EDELWEISS}
starting to probe interesting regions of parameter space (for an
extensive discussion see \cite{Berg}). The main problem currently lies
in the high level of background noise, either from ambient
radioactivity or cosmic-ray induced activity. Information on the
direction of the recoils could potentially also be useful and
yield a large improvement in sensitivity \cite{Spergel}.

In all these experiments, the count rate strongly depends on the
velocity distribution of the incident particles, and a modulation
effect due to the orbital motion of the Earth around the Sun is
expected \cite{Drukier}. In most cases, an isotropic Maxwellian
distribution has been assumed (e.g. ~\cite{Freese88,Bernabei98}),
although there are other examples in the recent literature, discussing
multivariate Gaussian distributions \cite{Ullio01,Green01,wyn01}.
Attempts at understanding the effect of substructure in the velocity
distribution of dark-matter particles have also been made
\cite{sikivie98,stiff01,hogan01}. This substructure would have its
origin in the different merger and accretion events the Galaxy should
have experienced over its lifetime \cite{hw99}. 

The progressive build up of dark halos through mergers and accretion
of smaller subunits implies that the latter will leave substructure
in the phase-space of the final object. This is because the
phase-space volume of the final object is much larger than that
initially available for each one of the objects independently. For
example, for a small satellite galaxy the initial phase-space volume
occupied by its particles is proportional to $(R^{\rm sat} V_c^{\rm
sat})^3 $, where $R^{\rm sat}$ is the size of the satellite, and
$V_c^{\rm sat}$ its circular velocity. The volume available to the
satellite particles after the merging is determined by their orbit,
and is a factor $(R^{\rm gal}/R^{\rm sat})^3 \times (V_c^{\rm
gal}/V_c^{\rm sat})^3$ larger, where $R^{\rm gal}$ is the size of the
final object and $V_c^{\rm gal}$ its circular velocity. Note that even
in the case of a major merger, where the mass is doubled, the
phase-space volume available is already 4 times larger \footnote{This
is because $M \propto R^3 \propto V^3$, and thus if $M_f = 2 M_i$ then $
R_f^3 V_f^3 \propto 4 R_i^3 V_i^3$}. The key question is whether this
substructure will be directly or indirectly observable. For example,
if there is a bound satellite going through the Solar neighbourhood at
the present day, it will dominate the flux of dark-matter particles on
Earth. The energy spectrum of these particles will be strongly peaked
around the orbital energy of the clump, perhaps giving a signal
similar to a delta function.  As we shall show in
Sec.\ref{sec:boundhalos} the fraction of mass in satellites which
could have survived the tidal field of the Galaxy by the present day
is less than $10^{-2}$ of total mass of the Galaxy, implying
that such a scenario is relatively unlikely. 

More realistic is to assume that the satellite halos that contribute
with mass to the Solar neighbourhood will be completely disrupted. The
particles freed from such satellites will tend to follow the initial
orbit of their progenitor, and eventually will fill a volume
comparable to the size of the orbit.  Because of the conservation of
phase-space density (Liouville's theorem), this implies that locally
they should have very similar velocities \footnote{If initially
$\Delta_x \Delta_v$ is the phase-space volume occupied by the
satellite, and if $\Delta_x'$ is its final volume, then $\Delta_v' =
\Delta_v \times \Delta_x'/\Delta_x$, where as discussed above,
$\Delta_x'$ is the volume given by the orbit, and is much larger than
the original volume of the satellite.}.  Thus one may expect to see
streams of particles going through the Solar neighbourhood, which had
their origin in the different merging events.  Such streams have
already been observed in the motions of nearby halo stars and in the
outer regions of the Galactic halo \cite{helmi99,ibata94}. Streams
manifest themselves as peaks in the velocity distribution function.
Clearly it is important to determine for the dark-matter particles in
the vicinity of the Sun whether this distribution function will be
dominated by a few of these peaks, or whether their number is so
large, that it will be close to Gaussian.

The best way to understand the expected properties of the Galactic
halo in the Solar neighbourhood is through high-resolution simulations
starting from appropriate cosmological initial conditions.  Analytic
modelling can provide insights into the processes that drive the
build-up of structure such as phase-mixing, or tidal
stripping. Nevertheless it needs to be complemented by cosmological
simulations, that provide the mass spectrum of the accreted halos,
their orbital parameters, their characteristic merging times, and the
detailed mixing of the material they deposit. The highly non-linear
character of the hierarchical build up of a galaxy like the Milky Way,
forces us to resort to numerical simulations to make realistic
predictions for its properties. Very high resolution simulations are
required to be able to resolve the substructures leftover from merging
events, since their density contrasts are expected to fade rather
quickly with time (as $t^3$ for sufficiently long timescales
\cite{hw99}).

The main goal of the present paper is to understand the phase-space
structure of a dark-matter halo. We wish to quantify the expected
amount of substructure and understand its effect on direct dark-matter
detection experiments.  Particular emphasis will be put on determining
the properties of the dark-matter distribution in the Solar
neighbourhood: its mass growth history, the spatial distribution and
the kinematics of particles in this region of the Galactic halo.  We
address these issues by scaling down a high-resolution simulation of
the formation of a cluster of galaxies in a $\Lambda$CDM cosmology to
a galactic size halo \cite{vrs01a}. 

\section{Methodology}

The simulations we analyse here were carried out using a parallel
tree-code ~\cite{vrs01b} on the Cray T3E at the Garching Computing
Centre of the Max Planck Society. These simulations were generated by
zooming in and re-simulating with higher resolution a particular
cluster and its surroundings formed in a cosmological simulation (as
in \cite{Tormen97}). The $\Lambda$CDM cosmological simulation has
parameters $\Omega_0 = 0.3$, $\Omega_\Lambda = 0.7$, $h= 0.7$ and
$\sigma_8 = 0.9$. The cluster selected is the second most massive
cluster in the simulation and has a virial mass of $8.4 \times 10^{14}
h^{-1}\sm$. The particles that end up in the final cluster of the
cosmological simulation and in its immediate surroundings (defined by
a comoving sphere of $70 h^{-1}$ Mpc radius) were traced back to their
Lagrangian region in the initial conditions for re-simulation. The
initial mass distribution between $21$ and $70 h^{-1}$ Mpc was
represented by $3 \times 10^6$ particles.  In the inner region, where
the original simulation had $2.2 \times 10^5$ particles, new initial
conditions were created for $4.5 \times 10^5$, $2 \times 10^6$, $1.3
\times 10^7$ and $6.6 \times 10^7$ particles, and small scale power
was also added onto this volume. The original force softening was also
decreased to obtain better spatial resolution.  All simulations were
run from very high redshift until $z=0$.

We can scale the simulated cluster to a ``Milky Way" halo by scaling
the circular velocity so that at its maximum it is equal to
220~\kms. The scaling factor obtained in this case is $\gamma=
v_c^{cl}/v_c^{MW} \sim 9.18$.  The virial radius of our simulated
``Milky Way" dark-matter halo is $r_{\rm vir} = 228$~kpc. The
justification for this simple scaling relies both on theoretical and
numerical results \cite{lc93,moore99,jing00}. For example,
high-resolution numerical simulations have shown that the overall
properties of galaxies and clusters of galaxies, such as their density
profiles, number of satellites and formation paths have overlapping
statistical distributions for the two types of objects.

\section{The phase-space structure of the galaxy}

In this section we shall focus on the properties of the dark-matter
distribution, with particular emphasis on the vicinity of the
``Sun". We are interested in which halos contribute matter to this
region of the Galactic halo, what were their initial properties, and
when were they accreted. We also investigate what is the present-day
spatial and velocity distribution of material in the ``Solar
neighbourhood", and how the direct detection experiments may be fine 
tuned to determine the nature of dark-matter.

\begin{figure*}
\includegraphics[height=21cm]{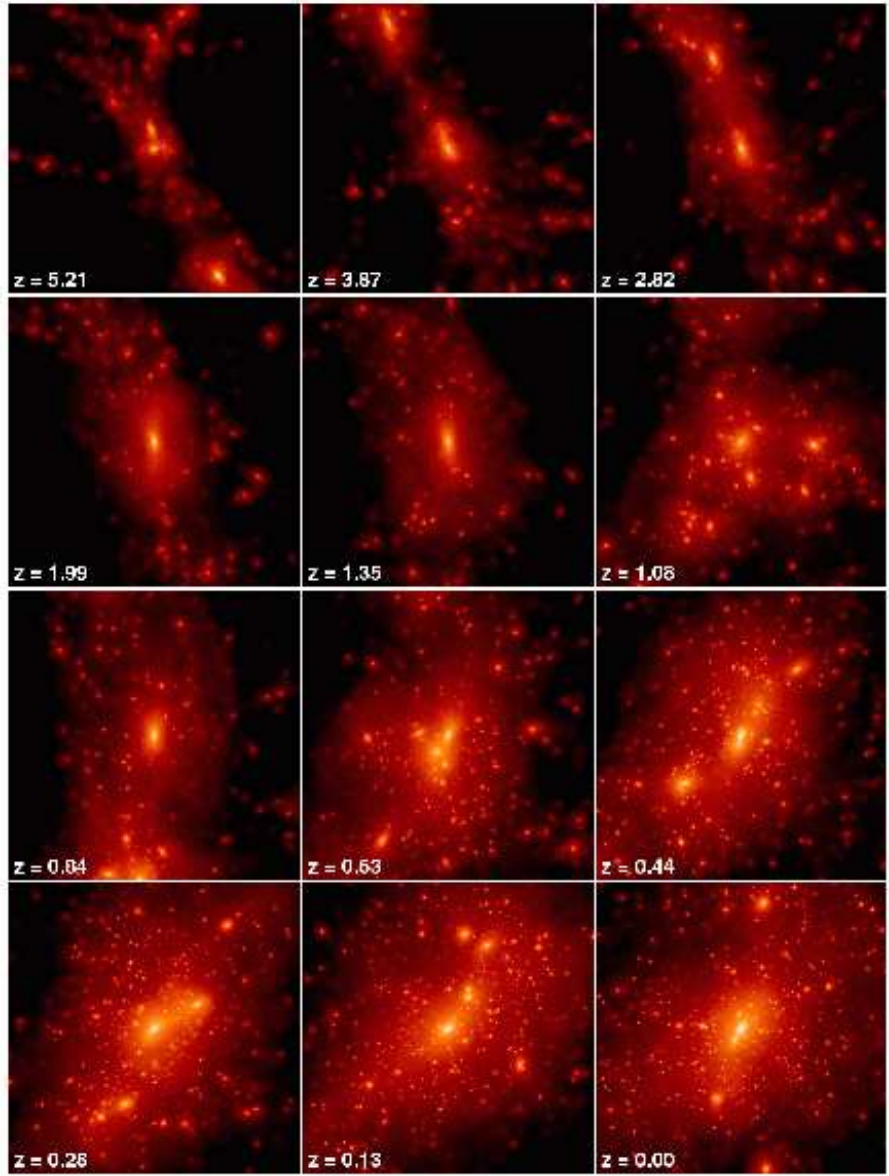}
\caption{Snapshots of the growth of the dark-matter halo in our
simulation. Each panel shows the projected mass density in a box of
side length 5.0 Mpc/h in the original cluster units, which correspond
to 778 kpc in the scaled units used throughout the paper. The panels
are centred on the main progenitor of the dark-matter halo at that
time.  The first panel corresponds to 12.7 Gyr ago, and the last panel
to the present time. Note how the halo grows through the merging and
accretion of smaller units.}
\label{fig:sims}
\end{figure*}
A series of snapshots of the growth of the cluster are shown in
Figure~\ref{fig:sims}. The simulations start from very small density
fluctuations, assumed to have been produced during the inflationary
expansion of the Universe \cite{Guth81}. Matter is then accreted onto
these initial density fluctuations through the action of gravity. A
dark halo forms when an overdense region decouples from the expanding
Universe, turns around and collapses onto itself. This process repeats
itself on progressively larger scales, and big halos are formed
through the merging and accretion of smaller units as shown in
Fig.~\ref{fig:sims}. These subunits will orbit the larger halo as
satellites for some time, as shown in the bottom panels in
Fig.~\ref{fig:sims}, until they are completely disrupted. We will
frequently refer to them as subhalos.  The progressive growth of mass
of the cluster is schematically shown in Figure~\ref{fig:merger_tree}
as a ``merger tree''.
\begin{figure}
\includegraphics[height=8.5cm]{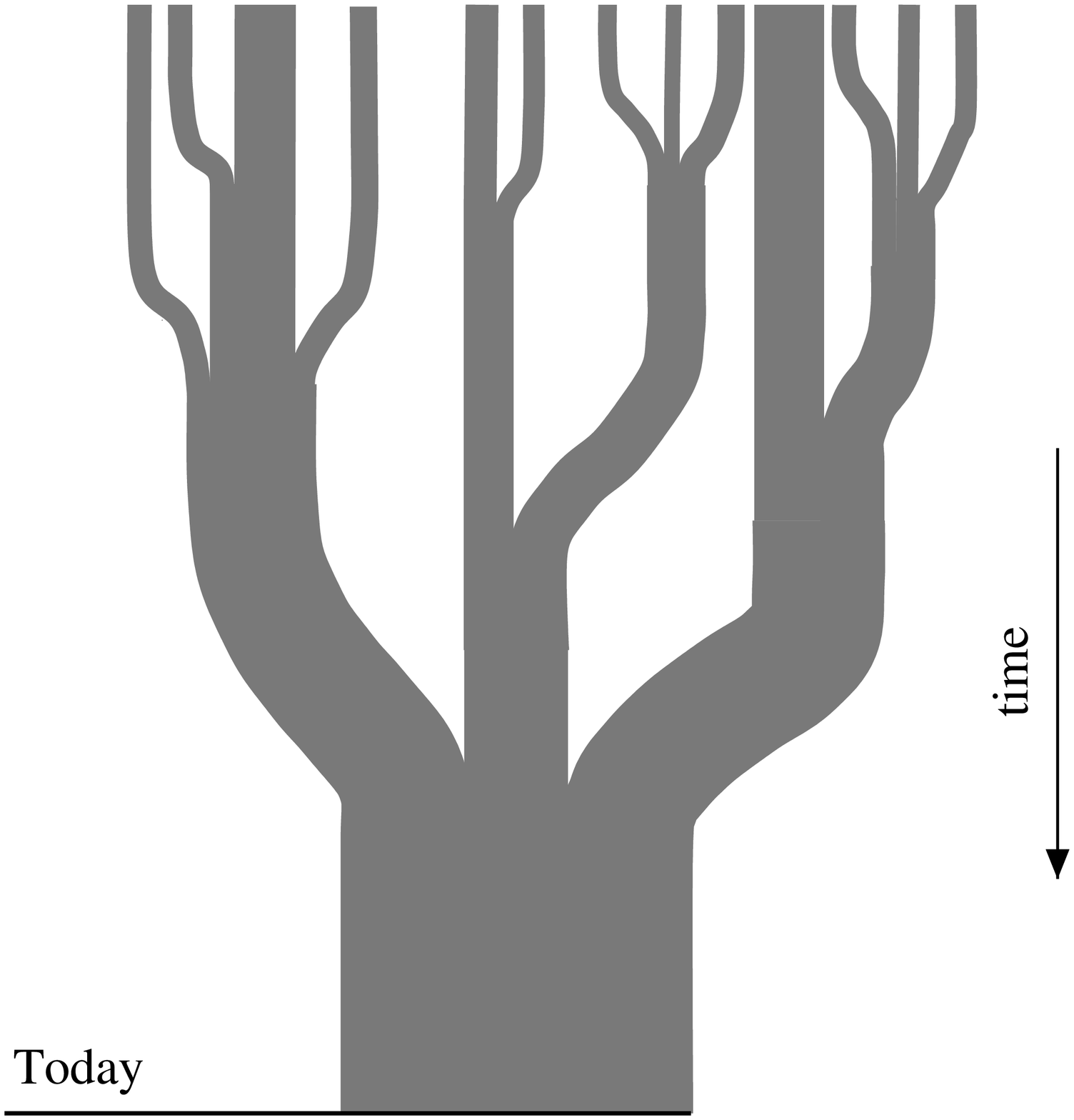}
\caption{Schematic representation of a ``merger tree'' showing the
growth of a halo in time. Time increases from top to bottom, and the
widths of the branches are proportional to the masses of the
progenitor halos (based on \cite{lc93}).}
\label{fig:merger_tree}
\end{figure}

\subsection{Mass growth history}

\subsubsection{The Galaxy}
Let us first study the properties of the dark-matter distribution as a
function of distance from the galaxy centre. We are interested in
determining what type of halos typically contribute to different
regions of the galaxy and their time of accretion.  This is relevant
because halos accreted at late times will be generally less mixed, and
could thus produce more massive streams dominating the velocity
distribution of particles near the Sun.  We also want to estimate the
probability that such halos could contribute to the Solar
neighbourhood mass budget.

We proceed by dividing the halo in six spherical shells around the
galaxy centre. These shells are located at: $ r < 10$ kpc, $10 \le r <
25$ kpc, $25 \le r < 50$ kpc, $50 \le r < 75$ kpc, $75 \le r < 100$
kpc and $100 \le r < 200$ kpc. For each particle in a shell, we
determine when it was accreted by the main progenitor of the galaxy
starting from redshift $z=2.4$ or 11 Gyr ago.  Particles may come from
accreted satellites or from the ``field". ``Field" particles are those
which did not belong to any bound structure before becoming part of
the galaxy. Because of our resolution limit, field particles may also
come from halos with less than 10 particles, i.e. with mass smaller
than $8.66 \times 10^5 \sm$.  The accretion time for particles in a
subhalo is defined to be the time of accretion of this subhalo. In
practice, we say that a subhalo $H_{\rm sub}$ identified at
redshift $z$ has been accreted by the main progenitor $H_{\rm main}$
at redshift $z'$, if at least half of the particles of $H_{\rm sub}$
are contained within $H_{\rm main}$ at $z'$, as well as the most bound
particle of $H_{\rm sub}$. For a particle from the field, the time of
accretion is simply defined as the earliest time at which this
particle has become a member of the main progenitor of the galaxy, as
determined by our FOF algorithm.

In Figure~\ref{fig:shells.m} we show the fraction of mass accreted
(normalised to the present mass) for each shell as a function of the
initial mass of the accreted satellite and for three different
redshift bins. We divide the analysis into the mass already present at
redshift $z =2.4$ (shown in dark grey); that accreted between $z =
2.4$ and $z=0.83$ (light grey); and between $z=0.83$ and the present
day (black).

The first panel of Figure~\ref{fig:shells.m} shows that the formation
time of the inner galaxy is strongly biased towards high redshifts,
with more than 60\% of the mass already present at $z=2.4$.  We also
note that late accretion does not play any role in building up the
inner galaxy. Subhalos with masses smaller than $10^7 \sm$ accreted at
late times do not make up more than $10^{-3}$ of the total mass in the
innermost shell.  This has implications for dark-matter detection
experiments, since it implies that recently accreted tiny subhalos
with very high phase-space density do not contribute enough to the
Solar neighbourhood to have a significant effect on the expected
signal, in contrast to the suggestion in \cite{moore01}. The only way
such small subhalos could make it to the vicinity of the Sun would be
by being accreted first by a large halo, which at some later redshift
has a major merger with the main progenitor of the Galaxy. We will
quantify how likely this may be in the next section.

Figure~\ref{fig:shells.m} also shows that small accreted subhalos tend
to deposit most of their mass at large distances. In these outer
regions, the contribution of heavy subhalos is of the same order of
magnitude as that from the smaller satellites.  The difference in the
final debris distribution from massive subhalos and from lighter ones
is due to dynamical friction; very massive satellites can sink to the
centre of the galaxy in short timescales, which enables them to
deposit a good fraction of their mass in the central regions. A large
fraction (about 20\%) of the mass in the outskirts comes from field
particles (as shown in the last panel of Fig.~\ref{fig:shells.m}),
strongly contrasting with the 0.7\% seen for the innermost shell.

Finally we remark that whereas the formation of the inner galaxy is
strongly skewed to high redshifts, the outer regions grow much more
gradually in time, with accretion still being important at late times.

\begin{figure}
\includegraphics[height=8cm]{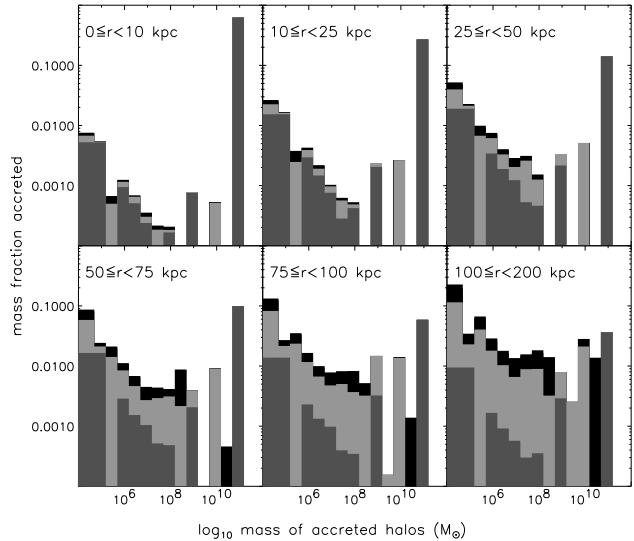}
\caption{The panels show the accreted mass fraction (normalised to the
present-day mass) as a function of the initial mass of the satellite
halo for spherical shells around the galaxy centre. The different
colours correspond to the fraction of mass accreted at different
redshifts. Dark grey corresponds to the mass already in place at
$z=2.4$; light grey to mass accreted between $z=2.4$ and $z=0.83$; and
black to that accreted between $z=0.83$ and the present day.}
\label{fig:shells.m}
\end{figure}

\subsubsection{The Solar neighbourhood}
\label{sec:SN_shells}

We now focus on the ``Solar neighbourhood", and analyse the region: 7
kpc $< r <$ 9 kpc. Proceeding as before, we determine the origin of
each particle in this spherical shell, and when it was added to the
galaxy halo. The growth of mass can take place through mergers or
accretion of subhalos, or through a smooth accretion due to the
progressive incorporation of field particles.

In the top panel of Figure \ref{fig:merger_time} we show the mass
accreted at a given time (to be more precise between two consecutive
outputs) $f_m(t)$ normalised to the total present-day mass in the
spherical shell 7 kpc $< r <$ 9 kpc. We also show the mass accreted
for the whole galaxy halo (thus for $r < r_{\rm vir}$) as a function
of time. We thus confirm that all the mergers that contributed a
substantial amount of mass to the ``Solar neighbourhood" took place
quite early. Mergers at late times contributed a relatively large
amount of mass to the galaxy halo, but deposited most of this mass in
the outer regions of the galaxy. 

Particles accreted in the past Gyr, do not account for more than about
$10^{-3}$ of the total present number of particles near the Sun. The
influence of streams from such recently accreted material on the
velocity distribution function near the Sun will thus be relatively
small, and may dominate only the high energy tail of the distribution
function.

\begin{figure}
\includegraphics[height=8.5cm]{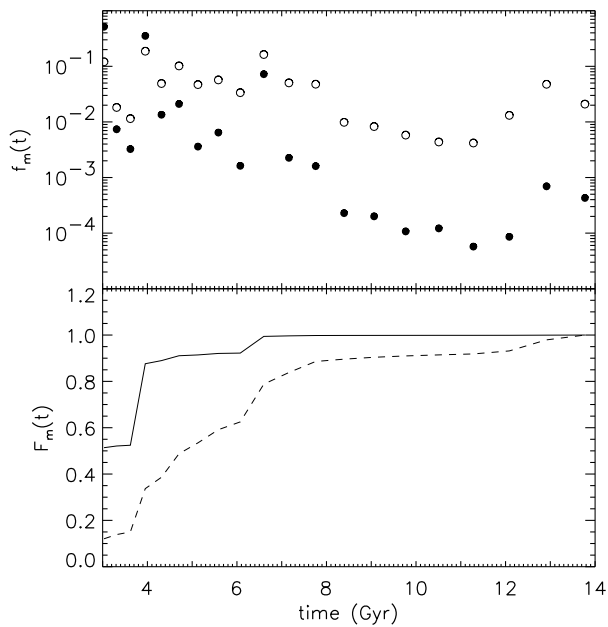}
\caption{Filled circles in the top panel show the growth of mass in
the ``Solar neighbourhood" normalised to the mass present at
$z=0$. Open circles correspond to the growth of mass of the whole
galaxy halo. The bottom panel shows to the cumulative growth of
mass. Note that 85\% of the mass in the Solar neighbourhood was
already in place 10 Gyr ago. For the whole galaxy halo (dashed curve),
this did not happen until 6 Gyr ago.}
\label{fig:merger_time}
\end{figure}

The lower panel in Figure \ref{fig:merger_time} shows the growth of
mass ($F_m(t) = \int_0^t f_m(t') {\rm d}t'$) as a function of time. We
see that more than 50\% of the mass that ends up in the ``Solar
neighbourhood" today was already in place 11 Gyr ago, and about 90\%
10 Gyr ago. Of course, this depends on the specific merger history of
this halo, since the large increase of mass observed at $t=4$ Gyr is
due to a major merger taking place at the time. However, we also
notice that after $t=7$ Gyr, there is almost no increase of mass in
the vicinity of the ``Sun". The large contrast with the gradual mass
growth of the halo as a whole can be clearly perceived from this
figure.

\subsection{Spatial distribution in the Solar neighbourhood}

One of the critical issues in understanding the outcome of the various
dark-matter experiments consists in characterising the expected
signal. As discussed in the introduction, of fundamental importance is
to know whether the distribution of particles in the vicinity of the
Sun is smooth or might be dominated by a just a few streams or even
bound lumps (e.g. \cite{moore01}).  Below we shall describe the
velocity distribution of particles that end up in this region of the
galaxy halo. Here we focus on their spatial properties.

In Figure \ref{fig:xyz_sun} we plot the positions of all particles
inside a cubic volume of 2 kpc on a side, located 8 kpc from the
``Milky Way" centre, which we assume is the distance between the Sun
and the Galactic centre. Because the dark halo is triaxial,
although almost prolate (the axes ratios are $I_1:I_2:I_3 =
0.65:0.71:1$), we assume the Galactic disk to be perpendicular to the
major axis of the halo.  The spatial distribution of particles inside
this representative volume is extremely smooth, as shown in Figure
\ref{fig:xyz_sun}.  This is mostly due to the fact that the material
that ends up in the inner galaxy mostly comes from a few very massive
halos. The very short dynamical timescales in this region of the
galaxy are also responsible for the very rapid and efficient mixing,
after which there remains very little or no spatial information on
their origin. 

\begin{figure}
\includegraphics[height=7cm]{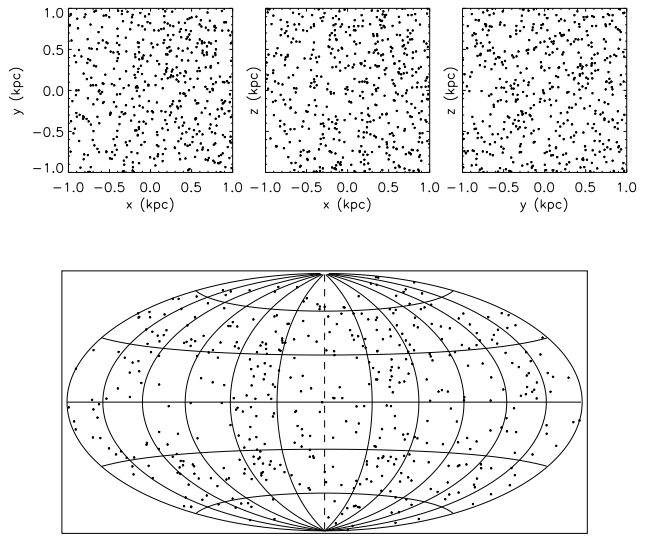}
\caption{The top panels show the spatial distribution of particles
inside a 2 kpc on a side volume located at 8 kpc from the galaxy
centre, i.e. this volume is centred on the ``Sun". There are 474
particles in this box. The bottom panel shows their distribution on
the sky.}
\label{fig:xyz_sun}
\end{figure}

\subsubsection{The properties of the halos that contribute to the Solar
neighbourhood}

To determine the characteristics of the halos that have contributed
matter to the Solar neighbourhood, we focus on the spherical shell: 7
kpc $ < r < 9$ kpc. As in Sec.\ref{sec:SN_shells}, we identify the
origin of the particles that are located in this shell at the present
time. We group the parent satellites according to their initial mass
in logarithmically spaced bins of width $d log M = 0.5$ starting from
$10^{5.5} \sm$ to $10^{12} \sm$. We estimate the contribution from
halos identified at three different redshifts.  Either such halos are
directly accreted by the galaxy, or they are accreted by another more
massive subhalo, which eventually merges with the galaxy.

In the left panel of Figure \ref{fig:mass_fraction_SN} we show the
contribution of matter to the spherical shell normalised to the
present mass in the shell. Thus we see that a large fraction of the
mass in this shell comes from the most massive halos identified at
$z=2.4$, and very little from field particles (as was also shown in
Figure~\ref{fig:shells.m}). For higher redshifts the largest
contribution tends to come from smaller satellites, which is
just a consequence of the fact that the heaviest halos have not yet
collapsed at these redshifts. For sufficiently high redshift, the
field particles (i.e. from unresolved halos) are the largest
contributor to the mass present today in the Solar neighbourhood.

Just a few individual halos contribute to the largest mass
bins. However, for the small mass bins, the number of halos
contributing within a given bin actually increases dramatically, as
shown in the right panel of Figure~\ref{fig:mass_fraction_SN}. The
large fraction of the mass in the spherical shell 7 kpc $ < r < 9$ kpc
that comes from halos with masses $M$ in the range $10^{5.5} - 10^6
\sm$ for $z \sim 10$, thus originates in a large number of small
independent halos.
\begin{figure*}
\includegraphics[height=8cm]{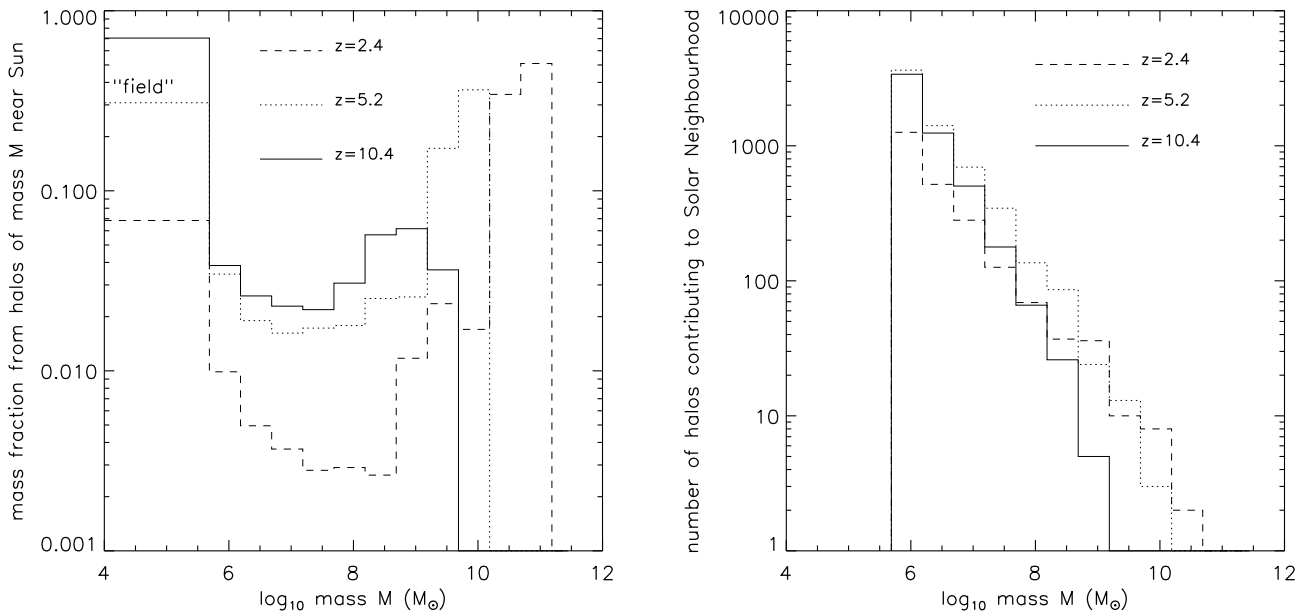}
\caption{The left panel shows the fraction of mass in a shell around
the ``solar" radius coming from halos identified at three different
redshifts, normalised to the present mass in the shell. The first bin
shows the contribution of field particles at the different redshifts,
which end up in the ``Solar neighbourhood". By $z=10.4$ not many
massive halos have yet collapsed, and almost all particles come from
the field. However, by $z=2.4$, most of these field particles are
found in halos. The largest contribution to mass in the vicinity of
the Sun comes from halos heavier than $10^{10} \sm$. The right panel
shows the number of contributing halos in the different mass bins.}
\label{fig:mass_fraction_SN}
\end{figure*}

\subsubsection{Are there any bound subhalos in the Solar
neighbourhood?}
\label{sec:boundhalos}

There have been recent suggestions in the literature that there could
be a population of very tiny subhalos orbiting in the Solar neighbourhood
\cite{moore01}. These subhalos which collapsed at very high redshift,
might have sufficiently large densities to survive almost intact
until the present day. If this picture were correct, their presence
would produce a signal on dark-matter detectors that would be very
different from that of a Gaussian distribution coming from a smooth
halo. 

Figure~\ref{fig:mass_fraction_SN} shows that the contribution of mass
from halos of $10^{5.5}- 10^6 \sm$ is not negligible, ranging from
0.9\% for halos identified at $z=2.4$ to 4\% for those identified at 
$z=10.4$. However, none of these halos has managed to survive bound in
our simulations, and so the mass they have contributed is in a smooth
component at the present time.

However we also need to quantify the probability that some of the mass
actually comes from bound subhalos below our resolution limit. To
tackle this problem we need to estimate the fraction of the mass in
such subhalos which remained bound until the present day with respect
to the total mass of the galaxy.  We shall do so using all subhalos
orbiting within the virial radius of the galaxy today. These subhalos
are mostly found in the outskirts of the galaxy, where their survival
times are longer due to the smaller galactic tidal forces.

The subhalo mass function $dN/dM$ gives the number of satellites of
the galaxy halo with mass in $[M, M+dM]$. With a sophisticated subhalo
finder it is possible to determine $dN/dM$ for our simulated halo at
the present time \cite{vrs01a}. Figure~\ref{fig:dNdM} shows that the
number of subhalos in a given mass bin can be well fit by a power law:
\begin{equation}
\frac{dN}{dM} = 1.45 \times 10^{-4} \frac{h}{\sm}
\left[\frac{M}{10^{7}\sm h^{-1}}\right]^{-1.73}.
\end{equation}
Although the total number of halos with masses smaller than $M$
diverges, the total mass in these halos is a well-defined and finite quantity
\[M_T(<M) = \int_{0}^{M} \frac{dN}{dM'}  M' dM'. \]
Thus
\begin{equation}
M_T(<M) = 5.29 \times 10^{10} \left[\frac{M}{10^{7}\sm h^{-1}}\right]^{0.27} 
\end{equation}
in $\sm h^{-1}$. Since the total mass of the galaxy halo in our
simulation is $M_{\rm host} = 1.2 \times 10^{12} \sm h^{-1}$, the
fraction of mass contained in subhalos of mass smaller than $M$ is
$p(<M) = M_T(<M)/M_{\rm host}$:
\begin{equation}
p(<M) = 4.4 \times 10^{-2} \left[\frac{M}{10^{7}\sm
h^{-1}}\right]^{0.27}.
\end{equation}
Thus for example, for the whole galaxy halo the fraction of mass in
bound satellites smaller than $10^{7} \sm h^{-1}$ at the present time
is $4.4 \times 10^{-2}$. This overestimates the fraction of mass in
subhalos orbiting the inner galaxy, since the spatial distribution of
subhalos is skewed towards large distances from the galaxy centre,
where tidal forces are weaker. For example, if we estimate $p(<M)$
only from subhalos within 30 kpc from the galaxy centre, we find that
the fraction of mass in bound satellites smaller than $10^{7} \sm
h^{-1}$ is $9.7 \times 10^{-3}$, a factor of 5 smaller than previously
found. We may conclude that at most about 1 out of one hundred events
detected in any dark-matter experiment are expected to come from
particles in subhalos with mass below $10^{7} \sm h^{-1}$.
\begin{figure}
\includegraphics[height=8cm]{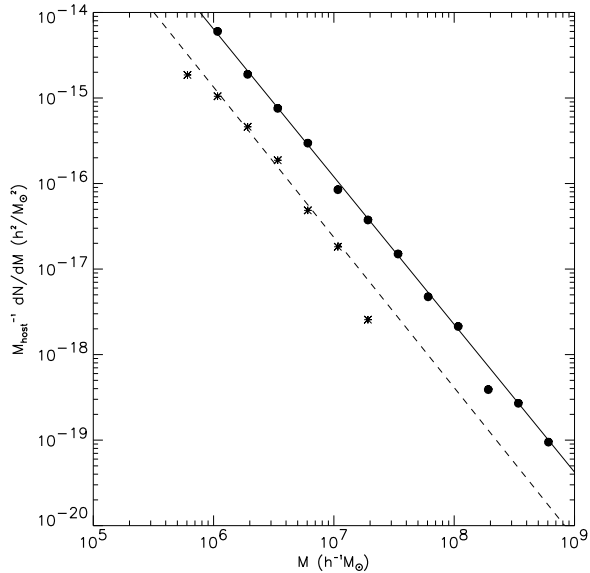}
\caption{Differential satellite mass function $M^{-1}_{\rm host}
dN/dM$ at redshift $z=0$ normalised to the mass of the host halo. The
solid circles correspond to the whole population of satellites, while
the asterisks only to those located in the inner 30 kpc. In the latter
case $M_{\rm host}$ is the mass within this spherical region.  The
straight line corresponds to $\log dN/dM = a + b \log M$, showing that
the differential mass function is very well fit by a power law.}
\label{fig:dNdM}
\end{figure}

\begin{figure*}[t]
\includegraphics[height=17cm,angle=90]{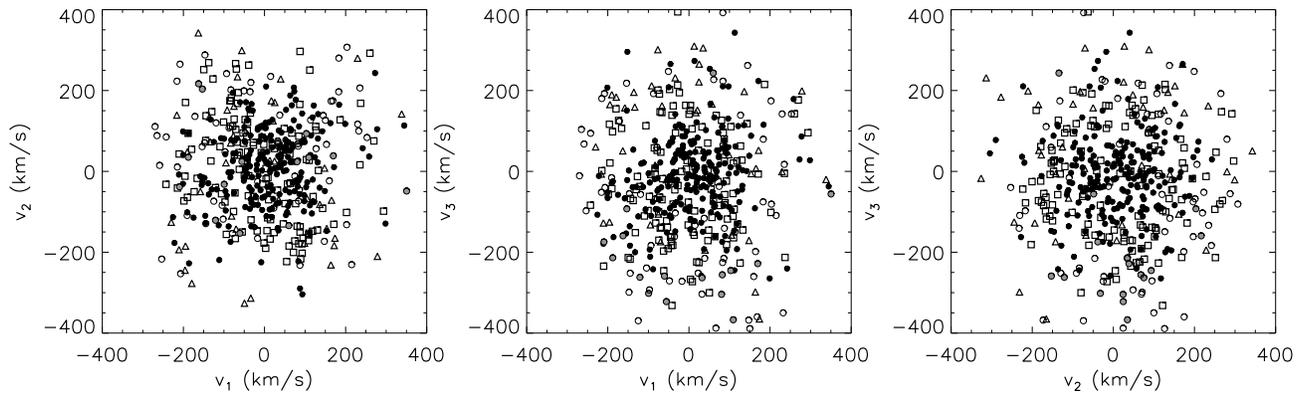}
\caption{Principal axes projections of the velocities of particles
located in a box of 2 kpc on a side on the ``Solar" circle, where all
quantities have been scaled to the ``Milky Way halo'' as described in
the text. There are 474 particles in this box. The different colours
and symbols are used here to indicate particles originating in the same halo
identified 11 Gyr ago. Open circles correspond to particles from halos
which do not contribute substantially to this volume. Black filled
circles are particles from the main progenitor identified at this
redshift (226 particles, i.e. 47\%). Open triangles correspond to
``field" particles, i.e. not associated to any halo 11 Gyr ago (42
particles, i.e. 9\%).  The squares (129 particles, i.e. 27\%) correspond to
the second most massive halo identified at this time, which merged
with the galaxy about 10 Gyr ago. The light grey circles (16
particles, 3\%) are for the third most massive halo, which merged with
the galaxy about 7 Gyr ago. These events are clearly visible in
Figure~\ref{fig:merger_time}, as having contributed most of the mass
in the ``Solar neighbourhood" in the history of the galaxy.}
\label{fig:boxes_col}
\end{figure*}
\begin{figure}[b]
\includegraphics[width=8cm]{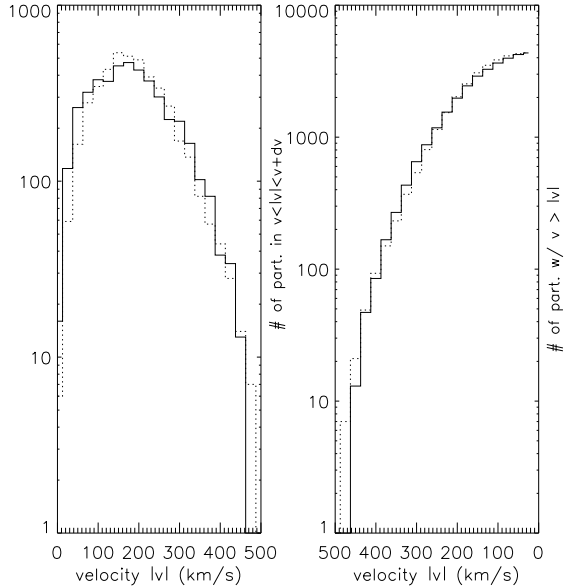}
\caption{For particles located in a box of 4 kpc on a side at the
``Solar'' radius, we plot the differential (left) and the cumulative
(right) velocity distributions (solid histograms). The dotted
histograms correspond to the expected distribution for a multivariate
Gaussian with the same velocity ellipsoid and number of points as the
data. The main differences are that the actual distribution of
velocities appears to be broader and with a sharper cutoff than the
Gaussian, and it is slightly less peaked.}
\label{fig:df_DM}
\end{figure}

\subsection{Kinematics in the Solar neighbourhood}

In Figure \ref{fig:boxes_col} we plot the velocities of particles
inside a cubic volume of 2 kpc on a side, located at 8 kpc from the
``Milky Way" centre. (The same volume as in Figure~\ref{fig:xyz_sun}.)
We identify with different colours and symbols particles that belonged
to the same halo at $z= 2.4$. At this redshift, which corresponds to
11 Gyr ago, we find 252003 halos with at least 10 particles in our
simulation.  In the box shown in this figure, there are 474 particles
which come from 39 different halos; only 3 contribute with more than
ten particles.  These three halos comprise the main progenitor of the
galaxy (i.e. the trunk in Figure~\ref{fig:merger_tree}) and the two
most massive halos that merged with the galaxy (see
Figure~\ref{fig:merger_time}.  We do not expect all particles of the
same colour to be clustered in a single massive stream, since each
individual halo is predicted to have given rise to many streams in the
Solar neighbourhood \cite{hw99,hws}. For example, the sixteen
particles originating in the third most massive halo (3\% of the
particles in this volume) are distributed in ten different
streams. Therefore, it is not surprising that it is difficult to
distinguish streams in this figure. The total number of particles
inside this box is too small to populate each expected stream with
more than one or two particles.

\subsubsection{The velocity distribution function}

Turning to a larger volume of 4 kpc on a side, allows us to increase
the number of particles by roughly a factor of 8.  In
Figure~\ref{fig:df_DM} we plot the velocity distribution function in
such a box in the vicinity of the ``Sun''. The left and right panels
show the differential and cumulative velocity distributions,
respectively.  We also show how these distributions compare to a
multivariate Gaussian with the same velocity ellipsoid as the
data. Although slight differences are visible, it is hard to
distinguish the two distributions from one another without making a
detailed statistical comparison.

\subsubsection{The fastest moving particles}

 In Figure~\ref{fig:DM_4kpc} we show the velocities of particles
located in the same 4 kpc on a side box of Fig.~\ref{fig:df_DM}. We
also note here that their velocity distribution is relatively smooth.
However, if we focus on the highest energy particles this seems no
longer to be the case, as shown by the particles highlighted in
grey. The 1\% fastest moving particles are strongly clumped.  Most of
this signal comes from a halo of mass $1.94 \times 10^{10} \sm$ that
merged with the galaxy at $z \sim 1$. Figure~\ref{fig:sky_DM} shows
the directions of motion of all particles in the box, where we again
highlight the fastest moving ones. Their distribution is clearly
anisotropic. 
\begin{figure*}
\includegraphics[height=5.2cm,width=17.5cm]{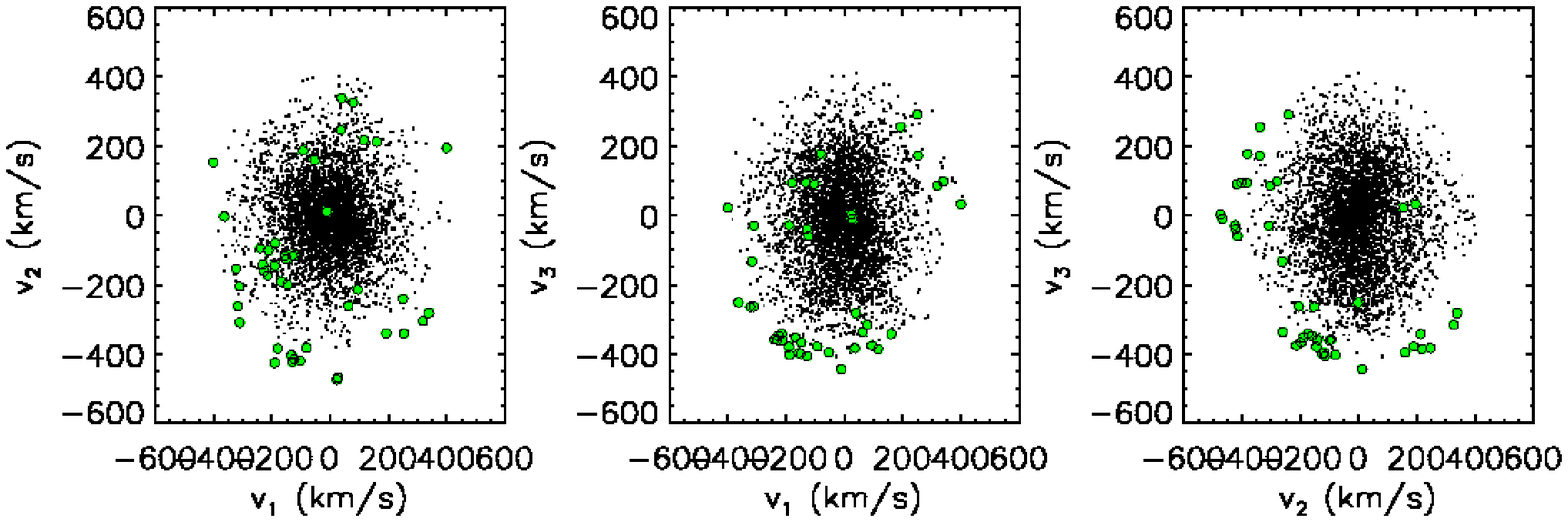}
\caption{Principal axes projections of the velocities of particles
located in the same box of Fig.~\ref{fig:df_DM}. Of the 4362 particles
present in this volume, we highlighted the 1\% fastest.  The velocity
dispersions along the principal axes are $\sigma_1 = 111.2$ \kms,
$\sigma_2 = 120.1$ \kms, and $\sigma_3 = 141.4$ \kms. The lump with
$v_1 \sim -185$ \kms, $v_2 \sim -140$ \kms and $v_3 \sim -370$ \kms
corresponds to a halo of $1.94 \times 10^{10} \sm$ identified at $z=
2.4$, and accreted at $z=1$, or $8.2$ Gyr ago.}
\label{fig:DM_4kpc}
\end{figure*}
\begin{figure}
\includegraphics[width=8.7cm]{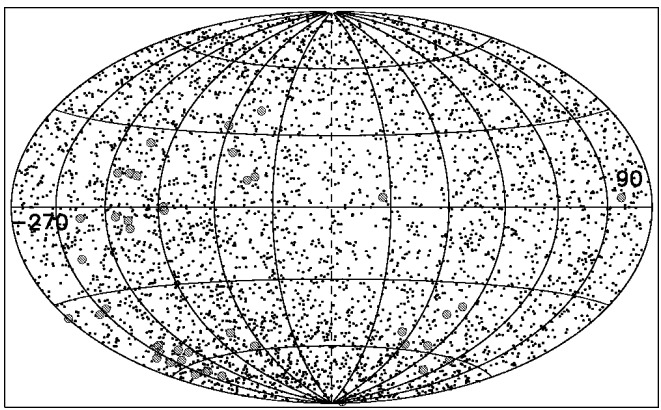}
\caption{This plot shows the directions of motion of the same
particles discussed in Figs.~\ref{fig:df_DM} and \ref{fig:DM_4kpc}. We
highlight in grey the 1\% fastest moving particles. The position of a
particle in the plot is given by the spherical angular coordinates of
its velocity vector, e.g. $v_1 = v \,{\rm cos}\phi \,{\rm cos}\theta$,
where $\theta$ is the latitude and $\phi$ the longitude.}
\label{fig:sky_DM}
\end{figure}
\begin{figure}
\includegraphics[height=8.5cm]{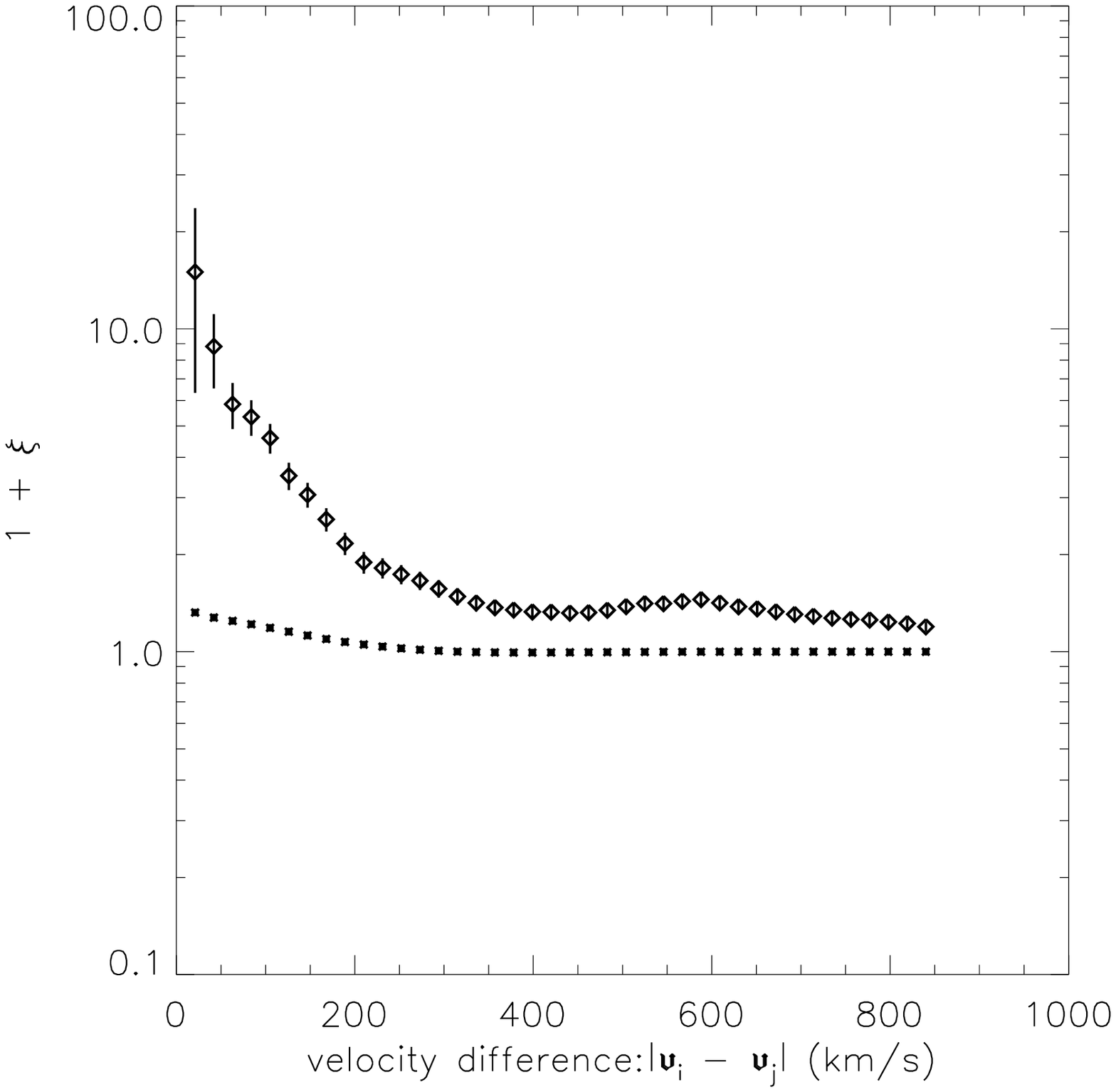}
\caption{For the same box as in Figures~\ref{fig:df_DM},
\ref{fig:DM_4kpc} and \ref{fig:sky_DM}, we plot the ``(averaged)
velocity correlation function" $\langle \xi \rangle$. It is defined as
the number of neighbours with velocity difference less than a given
value compared to what is expected for random deviates. In this case
the random deviates are drawn from a multivariate Gaussian with the
same number of particles and velocity dispersion tensor as the
data. Asterisks correspond to $\langle \xi \rangle$ over all the
particles in the box, whereas diamonds to the 1\% fastest moving
particles. Both for the full data set and for the fastest moving
particles there is a signal at small velocity differences, indicative
of the presence of streams. This signal is clearly much stronger for
the subset of most energetic particles. The error bars are based on
Poissonian counts.}
\label{fig:dd_DM}
\end{figure}

\subsubsection{The velocity correlation function}

To quantify the deviations from a smooth Gaussian distribution due to
the velocity substructure present in a volume in the Solar
neighbourhood, we compute the correlation function in velocity
space. We define the (averaged) velocity correlation function $\langle
\xi \rangle$ as
\begin{equation}
\langle \xi \rangle = \frac{\langle DD \rangle }{\langle RR \rangle } - 1
\label{eq:xi}
\end{equation}
(e.g. \cite{kerscher}) where $\langle DD \rangle$ is the number of
pairs of particles in our simulation with velocity difference less
than a given value, i.e.
\begin{equation}
\langle DD \rangle = \sum \mbox{pairs of particles $i,j$ with }  
 |{\bf v}_i -  {\bf v}_j| < \Delta.
\end{equation}
 $\langle RR \rangle$ is defined analogously for the same number of random
points. We estimate the error in $\langle \xi \rangle$ as
\begin{equation}
\Delta_{\langle \xi \rangle} = \frac{1 + \langle \xi
\rangle}{\sqrt{\langle DD \rangle }}.
\end{equation}

We compare the motions of particles located in the box of
Figures~\ref{fig:df_DM}, \ref{fig:DM_4kpc} and \ref{fig:sky_DM} with
those expected from a smooth Gaussian distribution. We generate
$N_{\rm real} = 10$ different Monte Carlo simulations with the same
velocity dispersion tensor and number of points as observed in this
box located in the vicinity of the Sun.  We compute $\langle \xi
\rangle$ as in Eq.~(\ref{eq:xi}), with
\begin{equation}
\langle RR \rangle = \frac{\sum_{i=1}^{N_{\rm real}} {\langle RR
\rangle}_i}{N_{\rm real}},
\end{equation}
where the sum is over the $N_{\rm real}$ different Monte Carlo
realizations. Thus $\langle RR \rangle$ is the number of pairs of
random deviates averaged over the ten Monte Carlo realizations.
Figure~\ref{fig:dd_DM} shows that there is a small, but statistically
significant, excess of particles with similar velocities (i.e. below
100~\kms) with respect to what would be expected for a multivariate
Gaussian distribution.  However if we focus on the 1\% fastest moving
particles, the excess of pairs of particles with similar velocities is
very noticeable, and is a clear indication of the streams visible in
Fig.~\ref{fig:DM_4kpc}.  Analysis performed with the 5\% fastest
moving particles still shows a significant deviation from a
multivariate Gaussian, albeit of smaller amplitude. For the 10\%
fastest moving particles, the deviation is as large as that observed
for the full data set, and would thus only be visible with a large
number of detection events, i.e. of at least a few thousand 
dark-matter particles.

Although the results presented here correspond to the analysis of just
one box in the ``Solar neighbourhood" the features observed here are
representative of what is seen for other similar volumes in this
region of the galaxy.

\begin{figure}
\includegraphics[height=8.3cm,angle=270]{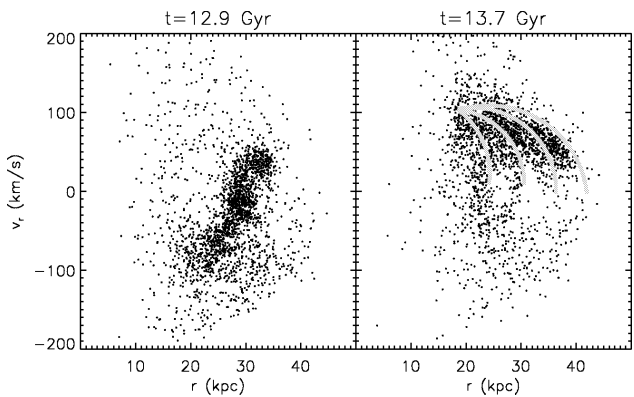}
\caption{Here we show the phase-space projection $(r,v_r)$ for
particles lost from a halo between $z=0.13$ and $z=0.06$ (between 12.1
and 12.9 Gyr). The left panel shows their distribution close to when
they were released. The right panel shows their present day
distribution. The multiple streams were produced in the lapse of at
most 1.7 Gyr. The grey curves have been added to highlight the
location of these streams in the diagram.}
\label{fig:streams}
\end{figure}

\section{Discussion}

The build up of dark-matter halos in a hierarchical universe is a very
nonlinear process, which happens through the merging and accretion of
smaller subunits.  The dominant dynamical processes at work are tidal
stripping, by which a satellite halo progressively loses its mass,
and phase-mixing by which this mass is progressively strung out along
streams. As an example we briefly discuss a halo of $4.3 \times
10^{10} \sm$ accreted at $z=1.8$ ($t=3.62$ Gyr), and the mass lost
between $z=0.13$ and $z=0.06$ ($t = 12.08$ and $t=12.91$ Gyr). In
Figure~\ref{fig:streams} we show the distribution of particles lost in
this redshift range, in a 2--dimensional projection $(r, v_r)$ of
phase-space. In the left panel we show their distribution just after
they were released from their parent satellite. The right panel shows
that what was initially a relatively coherent set of particles with
similar motions, has by the present time spread out into four
different streams which are seen to overlap in space. Note that this
happened in less than 1.7 Gyr, which shows how short the mixing
timescales are in the inner galaxy.

The density in a stream decreases in time, as the particles spread out
along the orbit of their progenitor system. Because the orbit defines
a 3--dimensional volume (a doughnut shaped volume defined by the
orbital turning points), the density in a stream will decrease as
$(t/t_{\rm orb})^{-3}$ \cite{hw99}. Thus, extrapolating the result
obtained for the material shown in Fig.~\ref{fig:streams} -- the
formation of 4 streams in 1.7 Gyr--, and noting the very early build
up of the galaxy halo (as demonstrated in Sec.~\ref{sec:SN_shells}),
we estimate that at least $4 \times (0.7 \, \mbox{Gyr}/0.35 \,
\mbox{Gyr})^3 \times (10 \, \mbox{Gyr}/1.7 \, \mbox{Gyr})^3 \sim 6500$
dark streams should be present in the Solar neighbourhood (a detailed
careful analysis, taking also into account other halos, shows that
this number should be even larger \cite{hws}). Here we used that the
orbital periods of particles in Fig.~\ref{fig:streams} are of the
order of 0.7 Gyr, while the median for particles near the Sun is
closer to 0.35 Gyr.

We note here that some authors \cite{hogan01,sikivie98} have assumed
that the density decrease in a stream is slower in time (linear or
quadratic), from which they (incorrectly) deduce that the streams
should have much higher densities. These works then argue that streams
can strongly affect the signal detected for the flux of dark-matter
particles going through the Earth.  We have just shown this is not the
case.

These results also apply to the density contrast in caustic surfaces
or rings, expected to form at the orbital turning points. We note here
that the proposed large effect associated with sheets of particles
falling in for the first, second or third time is not observed in our
simulations. This is {\it not} due to our finite resolution (as
suggested by \cite{sikivie01}), but due to the fact that the formation
time of the inner halo is so strongly biased to high redshifts. Since
particles which orbit the Solar neighbourhood have orbital periods of
the order of 0.35 Gyr, and the inner galaxy was already in place 10
Gyr ago, this implies that the density of these caustic structures
should be of the order of $(10 ~{\rm Gyr}/0.35 ~{\rm Gyr})^{-3} \sim 4
\times 10^{-5}$ times smaller today. For these reasons structures such
as caustic rings or surfaces will likely have no effect on the signal
that dark-matter experiments will detect.

Recent analytic work \cite{stiff01} has focused on the effect of
recently accreted material on the velocity distribution function near
the Sun. As these authors discuss, such material will mostly dominate
the high energy tail of the velocity distribution, and provide a
non-Gaussian and perhaps more easily detectable feature in the
spectrum of particles that go through the Earth. Although we reach a
similar conclusion in Sec.C.2, we do not agree on the magnitude of
this effect. In particular, we find that the density of streams from
such recently accreted material is a factor of ten smaller than has
been estimated in \cite{stiff01}. Probable causes for this discrepancy
can be perhaps be attributed to the different orbital distributions of
the accreted lumps (provided {\it ab initio} in our simulations, and
assumed to have a particular form in \cite{stiff01}); and in the
different methods used to measure the density of a stream (total
density of material from an accreted halo versus density of individual
streams).

Although our simulation represents the formation of a cluster halo,
rather than that of a galaxy, and it is only one possible realization
of the formation of a dark-matter halo, we believe the conclusions we
have reached are robust. For example, the mass growth history of this
particular halo is consistent with that of the Milky Way: most of the
mass was already in place in the cluster halo 10 Gyr ago, in good
agreement with the age of the oldest stars in the Milky Way disk.
High resolution simulations by other authors have shown that galaxy
and cluster sized halos grow in statistically similar ways and that
our particular halo is not unusual \cite{jing00} (see also
\cite{dzhao}).  Also, the properties of the velocity distribution of
dark-matter in the Solar Neighbourhood are in good agreement with the
observed kinematics of halo stars \cite{chibabeers}. Moreover, the
characteristics of the streams in our simulation, such as their low
densities and large number, are consistent with the simple analytic
estimates discussed above. Work along the lines of \cite{stiff01} can
provide further insight on how much variation one may expect to find
on a halo-to-halo basis.

\section{Conclusions} 

We analysed a high resolution simulation of the formation of a cluster
in a $\Lambda$CDM cosmology. By scaling it down to a galaxy size halo
(by the ratio of the maximum circular velocities) we were able to make
predictions for the expected dark-matter distribution near the Sun.

Our results indicate that direct detection experiments may quite
safely assume that the distribution of dark-matter particles in the
Solar neighbourhood is well represented by a multivariate Gaussian.
We find that none of the streams present in any of the volumes at the
Sun's distance from the Galactic centre dominate their local
distribution. The mean density of an individual stream is typically
0.3\% that of the local dark-matter distribution (deduced from the
number of particles in the rather dense stream shown in
Figure~\ref{fig:DM_4kpc}).  These small values are due to the fact
that most of the streams in the inner galaxy come from a few massive
halos that merged at high redshift to build up the object we see
today. These large halos mix extremely quickly and therefore give rise
to low density structures. Strong density enhancements such as those
predicted in \cite{sikivie98, sikivie01} are extremely unlikely in the
inner Galaxy.  Our simulation also shows that we should not expect to
find dense, recently formed streams near the Sun, since the last
accretion contributing matter to the Solar neighbourhood typically
took place about 1 Gyr ago, and provides only $\sim 10^{-4}$ of the
total mass in this region. Moreover, we find that fewer than 1\% of
the local dark-matter particles could be part of small dense subhalos
which have survived intact within the larger halo of the Milky Way. It
is therefore unlikely that an individual halo with these
characteristics will dominate the signal in direct detection
experiments.

Direct detection experiments which are sensitive to the direction of
motion of the fastest moving dark-matter particles may discover a
direct indication of the hierarchical growth of our Galaxy's halo. The
expected signal for these fastest moving particles is highly
anisotropic, and could be eventually be used, not just to determine
the nature of the dark-matter, but also to recover, at least
partially, the recent merging history of the Milky~Way.

We thank Ben Moore and Leo Stodolsky for useful discussions which
triggered many of the ideas discussed here; Uros Seljak and Anne
Green for helping us improve the manuscript. AH wishes to thank Joke
for being a continuous source of inspiration.


\newpage
\begin{references}
\bibitem{zwicky} F. Zwicky, Helv. Phys. Acta {\bf 6} 110 (1933). 
\bibitem{smith} S. Smith, \apj {\bf 83} 23 (1936).
\bibitem{rubin70} V. Rubin \& W.K. Ford, Astrophys. J. {\bf 159} 379 (1970).
\bibitem{faber79} S. Faber \& J.S. Gallagher, Ann. Rev. Astron. and
                     Astroph. {\bf 17} 135 (1979).
\bibitem{rubin80} V. Rubin, W.K. Ford \& N. Thonnard,
Astrophys. J. {\bf 238} 471 (1980).
\bibitem{peebles74} P.J.E. Peebles, Astrophys. J. Lett. {\bf 189} 51 (1974).
\bibitem{peebles82} P.J.E. Peebles, Astrophys. J. Lett. {\bf 263} 1 (1982).
\bibitem{blumenthal84} G.R. Blumenthal, S.M. Faber, J.R. Primack \&
M.J. Rees, Nature {\bf 311} 517 (1984).
\bibitem{frenk83} M. Davis, G. Efstathiou, C.S. Frenk \& S.D.M. White,
\apj {\bf 292} 371 (1985).
\bibitem{PC} R. Peccei \& H.R. Quinn, Phys. Rev. Lett. {\bf 38} 1440
(1977).
\bibitem{ax1} C. Hagmann {\it et al.}, Phys. Rev. Lett. {\bf 80} 2043 (1998).
\bibitem{ax2} I. Ogawa, S. Matsuki \& K. Yamamoto, Phys. Rev. D {\bf 53}
1740 (1996).
\bibitem{DAMA00} R.~Bernabei {\it et al.}  [DAMA Collaboration],
Phys. Lett. B {\bf 480}, 23 (2000).

\bibitem{CDMS00} R.~Abusaidi {\it et al.}  [CDMS Collaboration],
Phys. Rev. Lett. {\bf 84}, 5699 (2000).

\bibitem{EDELWEISS} A.~Benoit {\it et al.}  [EDELWEISS Collaboration],
Phys.Lett. B {\bf 513} 15 (2001).

\bibitem{Berg} L. Bergstr\"om, Rep. Prog. Phys. {\bf 63} 793 (2000).

\bibitem{Spergel} D.N. Spergel, Phys. Rev. D {\bf 37} 1353 (1988).

\bibitem{Drukier} A.K. Drukier, K. Freese \& D.N. Spergel,
Phys. Rev. D {\bf 33} 3495 (1986).

\bibitem{Freese88} K. Freese, J. Frieman \& A. Gould, Phys. Rev. D {\bf
37} 3388 (1988).

\bibitem{Bernabei98} R.~Bernabei {\it et al.}  [DAMA Collaboration],
 Phys. Lett. B {\bf 424} 195 (1998).

\bibitem{Ullio01} P. Ullio \& M. Kamionkowski, J. High Energy
Phys. {\bf 103} 049 (2001). 

\bibitem{Green01} A.M. Green, Phys. Rev. D {\bf 63} 043005 (2001)

\bibitem{wyn01} N. W. Evans, C. M. Carollo \& P. T. de Zeeuw,
Mon. Not. R. Astron. Soc.  {\bf 318} 1131 (2000).

\bibitem{sikivie98} P. Sikivie, Phys. Lett. B {\bf 432} 139 (1998).

\bibitem{stiff01} D. Stiff, L.M. Widrow \& J. Frieman, Phys. Rev. D
{\bf 64} 083516 (2001).

\bibitem{hogan01} C. Hogan, Phys.Rev. D {\bf 64} 063515 (2001).

\bibitem{hw99} A. Helmi \& S. D. M. White, Mon. Not. R. Astron. Soc. 
{\bf 307} 495 (1999).

\bibitem{helmi99} A. Helmi, S. D. M. White, P. T. de Zeeuw \&
H.S. Zhao, Nature {\bf 402} 53 (1999).

\bibitem{ibata94} R. Ibata, G. Gilmore \& M. Irwin, Nature {\bf 370} 194
(1994).

\bibitem{vrs01a} V. Springel, S.D.M. White, G. Tormen \& G. Kauffmann,
 Mon. Not. R. Astron. Soc. {\bf 328} 726 (2001).

\bibitem{vrs01b} V. Springel, N. Yoshida \& S.D.M. White, New
Astron. {\bf 6} 79 (2001).

\bibitem{Tormen97} G. Tormen, F.R. Bouchet \& S.D.M. White,
 Mon. Not. R. Astron. Soc. {\bf 286} 865 (1997).

\bibitem{Guth81} A. Guth, Phys. Rev. D {\bf 23} 347 (1981).

\bibitem{lc93} C. Lacey \& S. Cole,  Mon. Not. R. Astron. Soc. {\bf
262} 627 (1993).

\bibitem{moore99} B. Moore, S. Ghigna, F. Governato, G. Lake,
        T. Quinn, J. Stadel \& P. Tozzi, \apj Lett. {\bf 524} 19 (1999).

\bibitem{jing00} Y.P. Jing, Y. Suto, \apj Lett. {\bf 529} 69 (2000).

\bibitem{moore01} B. Moore, C. Calcaneo-Roldan, J. Stadel, T. Quinn,
G. Lake, S. Ghigna \& F. Governato, Phys. Rev. D {\bf 64} 063508 (2001).

\bibitem{sikivie97} P. Sikivie, I.I. Tkachev \& Y. Wang, \prd {\bf 56}
1863 (1997).

\bibitem{freese01} K. Freese, P. Gondolo \& L. Stodolsky, \prd {\bf
64} 123502 (2001).

\bibitem{kerscher} M. Kerscher, I. Szapudi \& A. Szalay, \apj
Lett. {\bf 535} 13 (2000).

\bibitem{hws} A. Helmi, S. D. M. White \& V. Springel, in preparation.

\bibitem{sikivie01} P. Sikivie, astro-ph/0109296 (2001).

\bibitem{chibabeers} M. Chiba \& T. C. Beers, Astron. J. {\bf 119} 2843
(2000).

\bibitem{dzhao} D. Zhao, H. Mo, Y.P. Jing \& G. Boerner,
astro-ph/0204108 (2002).

\end{references}
\end{document}